# SCIENCE, TECHNOLOGY AND SOCIETY

What can be done to make science more appealing and easier to understand

G. GIACOMELLI AND R. GIACOMELLI

*Physics Department, University of Bologna and INFN, Sezione di Bologna*

We shall discuss some aspects of science and technology, their increasing role in the society, the fast advances in modern science, the apparent decrease of interest of the young generation in basic sciences, the importance of proper science popularization for better public education and awareness in scientific fields.

## 1. Introduction

Science is interested in the laws of nature, while Technology applies scientific knowledge to make new things, new machinery and it may be used to "dominate" nature and to improve our life. The two aspects are deeply connected: without scientific research there is no technological progress and without technology we would not have new instruments for research. Usually the technological research improves and creates new instruments in known scientific fields, while most of the great technological revolutions are spin off of fundamental research. Just one example: WWW (World Wide Web), the key which opens every gate of Internet, the prefix most used by web navigators, was invented for improving communication in fundamental research in a large European Laboratory for fundamental physics, CERN in Geneva.

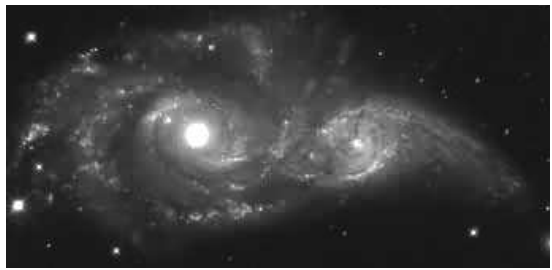

Fig. 1. Two colliding galaxies. They are not a galaxy and an antigalaxy because from the colliding region at the center of the picture we do not observe any drastic increase of luminosity, as it would be expected from particle-antiparticle annihilations.

Why do we perform research in general, and in particle physics in particular? For particle physics the standard answers are of the following type:

i) to understand the structure of matter and of what holds it together (Fig. 1), ii) to satisfy our curiosity, iii) because we enjoy doing it, iv) for technical spin-offs, v) for more modern teaching, others.

Can we justify the high costs of large scale research, like those in our field? It is now necessary to explain to the public what we do and how we spend the taxpayer money, besides assuring that we are not spoiling the environment. This is true in particular for research performed in the large international and national laboratories, like Brookhaven, CERN, etc. Scientific outreach (including science popularization, scientific awareness and appreciation of current research) has become an essential task for the research community and these activities must be made in a professional way [1,2].

Technological development leads to economic progress, increased well-being, to new medical applications and more. The applications coming from modern physics have changed communications (TV, cellular phones, Internet) and make it possible to look inside the human body without opening it (x-rays, nuclear magnetic resonance, ultrasound, Positron Electron Tomography, new computer applications, etc.). Biotechnology is changing and will change biology, medicine and our lives even more. But we may have to face also ethical problems.

The recent scientific and technological progress has not been accompanied by appropriate penetration in modern society of the fundamental scientific concepts, and in the media often there is a considerable amount of "parascience" and even of antiscience [1,2].

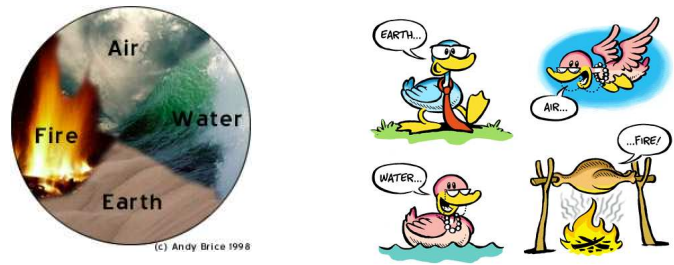

Fig. 2. **What is the World Made of?** (Particle Data Group). Why do so many things in this world share the same characteristics? People have come to realize that the matter of the world is made from a few fundamental building blocks of nature. By fundamental building blocks we mean objects that are simple and structureless -- not made of anything smaller. Since time immemorial people have been trying to understand what the Universe is made of. One of the earliest theories said that everything could be built from just four elements, Earth, Air, Fire and Water. This was a great scientific theory because it was simple. But it had one big drawback: it was wrong. The Greeks added also a fifth substance, quintessence, a term now used to describe the energy of the vacuum.

Science and technology play an increasing role in everyday life and progress in modern science and technology occurs quickly, both in specific subjects and because of the opening up of new fields and new interests. We should consider these changes in our school curricula, promote refresher courses, permanent education, and proper scientific outreach.

**2. Science popularization**

For researchers it is not easy to do science popularization and outreach in a simple and appealing manner, and one may remember what Gianni Puppi used to say in Bologna: "If you are not able to explain to your aunt in less than 5 minutes what you are doing in physics, then you have not really understood what you do", Fig. 2 [3].

Scientific knowledge update is needed to understand the great scientific and technological changes. But this must be done properly, stimulating the interest; the mass media often do not help since they may insist on aspects which are doubtful or incorrect or not scientific. An example: the press is full of articles about the "hydrogen economy", but they often forget to write that there are no "hydrogen mines"; they also write about getting hydrogen from water, forgetting that water is the result of combustion (it is like ash) and that to break water molecules into hydrogen and oxygen requires energy (which at the moment could only come from fossil fuels or nuclear fuels). A second example: vaccinations saved millions of people, but only the few unhappily lost lives become news in the press. Moreover the press and TV give great emphasis to magic, horoscopes, divinations and not enough to proper scientific information.

Most of the high school teachers followed university courses several years ago, when the teaching did not involve many of the present basic concepts, like subnuclear physics, the quark model, neutrinos, etc. It is clear that they would benefit from regular refresher courses.

While in the past the experimental research was done by small groups, now it is done by large collaborations involving many groups from different countries and large detectors which may be very costly [4,5]. Furthermore the decision on which experiments to perform may involve not only scientists, but larger communities, including bureaucrats, politicians, citizens. This means that the science communication should involve a broader community, using all available media, written journals, conferences, TV, Internet, etc.

Both in developed and developing countries there has been a consistent decrease of university applications in basic scientific fields (physics, chemistry, mathematics, ....). It seems that in the young generations there is a decreasing interest in sciences, even if the students, the teachers and the people would like to know more about science.

The Universities worry about it and they ask their scientists to increase their efforts for scientific outreach, in newspapers, TV, conferences, debates and by innovative means.

Also for developing countries there are many initiatives, see [6]. It may be worth recalling the following statement by a Peruvian officer: "Peruvian society is not well informed about science and people think that it is something too difficult to be understood by lay people. We need to change this scenario in order to promote debate in science and technology issues, using as many different tools as possible" [6].

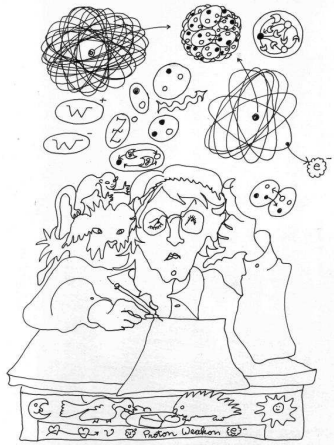

Fig. 3. My aunt after my explanation of particle physics (P. Waloscheck, DESY) [3].

## 3. Scientific outreach in Internet

The web site in Internet is one of the new methods of scientific outreach. All international and national large laboratories like CERN, NASA, ESA, INFN, have prepared nice outreach sites [7]. Now this is also done by smaller labs and by universities. Efforts are made to create simple and stimulating web sites, using interesting new approaches, with nice figures, often animated or interactive.

*The multidisciplinary outreach web site of the University of Bologna* is mainly addressed to last year high school students and university students, but it is also for a wider audience [8]. The purpose is to provide and promote scientific popularization and outreach in simple and appropriate ways, stimulating at the same time the curiosity of the younger generation. The site is "multidisciplinary": with topics such as The Brain, Antimatter, Molecular machines, Electrosmog, The radio window on the cosmos, Pollution of the environment, Artificial intelligence, Dark matter, Physics and fantasy, etc. Every page has a minimal text and uses photographs, figures, animated figures, and whatever can make the reading more inviting, without losing scientific correctness. Particular notes are also devoted to scientists from Bologna: Luigi Galvani (Animal electricity), Guglielmo Marconi (Wireless telegraphy), etc.

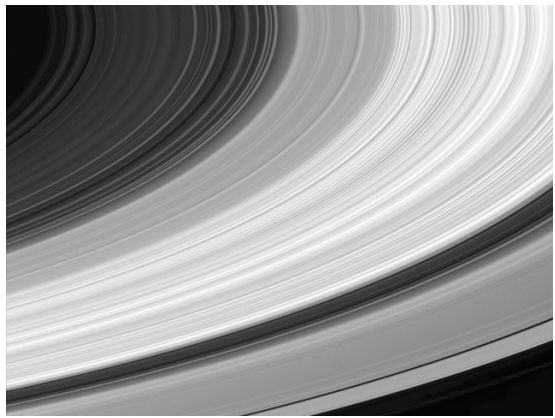

Fig. 4. Astronomy picture of the day (23/7/04): Saturn's Rings. Astronomy, astrophysics and space science pictures provide an incredible amount of beautiful pictures, which can easily be used for science popularization.

In order to facilitate the comprehension of technical terms an interactive dictionary is readily available on line for each subject.

There are links to science popularization journals, to scientific web sites, with the purpose to help students to widen their interests. The site contains also "articoli di approfondimento".

An example of the content of the topic *Antimatter*, written in cooperation with CERN, is:

Antimatter: what is it? Something exotic and not real?
    Everything you wanted to know about antimatter
Short history of antimatter
    From the first revolutionary ideas to the present situation
Antimatter around us
    From antimatter in cosmic rays to antimatter use in the PET
Antimatter at the beginning of the Universe
    How much antimatter was there? How did it disappear?

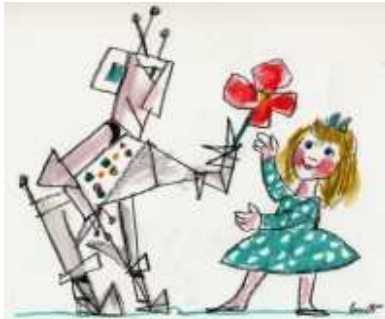

Fig. 5. Artificial intelligence. (Credit: "*Roboethics Symposium*"). Artificial intelligence is an interdisciplinary scientific and technological field in which studies are made on intelligence, logic, robotics, learning, theoretical and applied informatics [8].

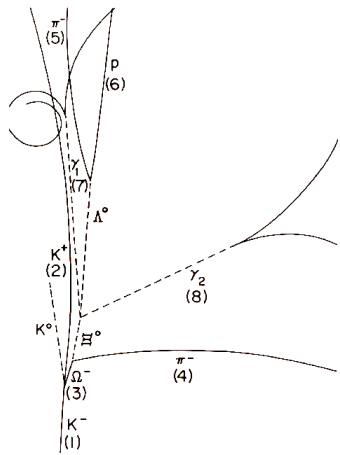

Fig. 6. The first observation of the Omega-minus particle in the Brookhaven hydrogen bubble chamber (1954). Selected bubble chamber photographs are very useful for the popularization of particle physics [3].

In the section "Antimatter around us", practical examples are given using cosmic rays, radioactive decays and their use in Positron Electron Tomography (PET). Fig. 1 is a figure from the section on "Antimatter today in the Universe". Examples of other figures used in other topics discussed in the site are Figures 5, 6, 7, 8.

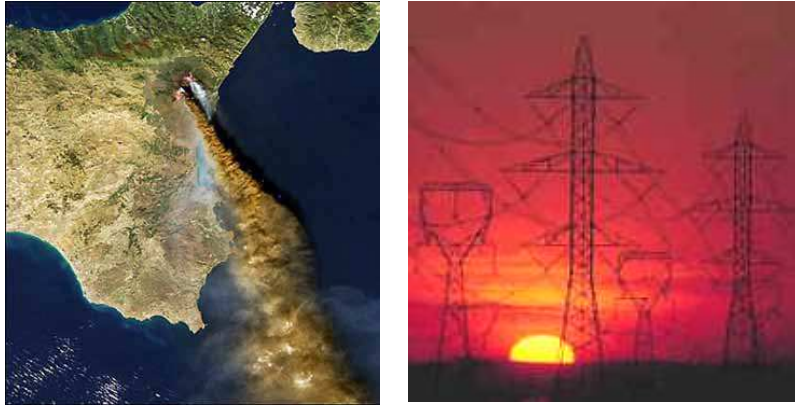

Fig.7 (a). Satellite photo of the Etna eruption in July 2002. (Credit: *Kathy Strabala and NASA*) [7, 8]; (b). Electrosmog (Electromagnetic pollution) is a term which generates wide discussions, often with exaggerated fears [8].

## 4. Conclusions

Science and technology play an increasing role in our lives, and progress in modern science and technology occur very quickly. Science and technology cannot give an answer to everything, but they lead to civic and economic evolutions improving the quality of our lives.

It is generally agreed that education and awareness in science have to be strengthened. Scientific outreach, improvements in teaching, proper scientific information are very important issues. Outreach should also be addressed to politicians and decision makers.

While for many researchers the main motivation for doing basic research remains scientific curiosity, for most of people the motivations involve also scientific progress, technological improvements, well being and the quality of everyday life, without spoiling the environment.

We should remember that the advanced techniques used in particle physics, can be applied in many fields, in particular in medicine.

We would like to thank A. Casoni, P. Catapano, B. Poli and many other colleagues for their cooperation.

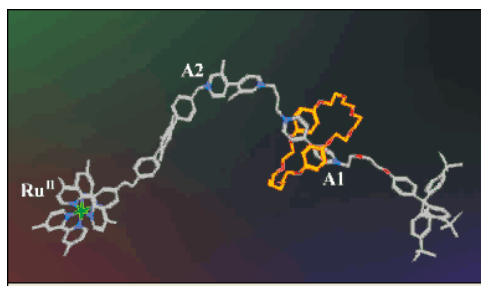

Fig. 8. Molecular machines. In order to reach extreme miniaturization, chemists start from molecules [bottom up approach]; assemblying them in appropriate ways they obtain "objects", "machines" (complicates molecules) of nanometer dimensions ($10^{-9}$m). The figure shows a "molecular shuttle" activated by luminous energy, see [8].